\documentclass[twocolumn,showpacs,amsmath,prl,aps,amssymb,superscriptaddress]{revtex4-1} 
\usepackage[T1]{fontenc}
\usepackage[latin9]{inputenc}
\usepackage{times}
\usepackage{color} 
\usepackage{xspace}
\usepackage{amssymb,amsmath}
\usepackage{amsbsy}
\usepackage[pdftex]{graphicx}
\usepackage{bm}
\usepackage{float}
\usepackage[normalem]{ulem}

\usepackage[unicode,breaklinks]{hyperref}
\hypersetup{
    unicode=true,
    plainpages=false, 
    colorlinks=true,
    linkcolor=blue,
    citecolor=blue,
    filecolor=black,
    urlcolor=blue
}
\usepackage{url}
\usepackage{verbatim}

\newcommand{\add}{a_{dd}}
\newcommand{\edd}{\epsilon_{dd}}
\newcommand{\br}{\mathbf{r}}
\newcommand{\bx}{\mathbf{x}}
\newcommand{\LGP}{\mathcal{L}_\mathrm{GP}}
\newcommand{\gammaQF}{\gamma_{\mathrm{QF}}}
\newcommand{\npeak}{n_{\mathrm{peak}}}
\DeclareMathOperator{\Ei}{Ei}

\begin{document}
 
\title{Collective excitations of self-bound droplets of a dipolar quantum fluid}

\author{D.~Baillie}  
\affiliation{Department of Physics, Centre for Quantum Science,
and Dodd-Walls Centre for Photonic and Quantum Technologies, University of Otago, Dunedin, New Zealand}
 \author{R.~M.~Wilson}   
\affiliation{Department of Physics, The United States Naval Academy, Annapolis, MD 21402, USA}
\author{P.~B.~Blakie}   
\affiliation{Department of Physics, Centre for Quantum Science,
and Dodd-Walls Centre for Photonic and Quantum Technologies, University of Otago, Dunedin, New Zealand}

\begin{abstract}  
We calculate the collective excitations of a dipolar Bose-Einstein condensate in the regime where it self-binds into droplets stabilized by quantum fluctuations. We show that the filament-shaped droplets act as a quasi-one-dimensional waveguide along which low angular momentum phonons propagate. The evaporation (unbinding) threshold occurring as the atom number $N$ is reduced to the critical value $N_c$ is associated with  a monopole-like excitation going soft as $\epsilon_0\!\sim\!(N-N_c)^{1/4}$. Considering the system in the presence of a trapping potential, we quantify the crossover from a trap-bound condensate to a self-bound droplet.  
 \end{abstract} 

\maketitle
 
Dipolar condensates consist of atoms with appreciable magnetic dipole moments that interact with a long-ranged and anisotropic dipole-dipole interaction (DDI).  
Recent experiments with dipolar condensates of dysprosium \cite{Kadau2016a,Ferrier-Barbut2016a,Schmitt2016a} and erbium \cite{Chomaz2016a} atoms have observed the formation of self-bound droplets that can preserve their form, even in the absence of any external confinement. These droplets occur in the dipole-dominated regime, where the DDIs dominate over short-ranged ($s$-wave) interactions, and for sufficiently many atoms in the droplet \cite{Baillie2016b,Wachtler2016b}. In the dipole-dominated regime meanfield theory predicts that the condensate is unstable to collapse, but as collapse begins and the density increases the (beyond meanfield) quantum fluctuation corrections become important. These Lee-Huang-Yang (LHY) \cite{LHY1957} corrections  \cite{Lima2011a,Lima2012a,Schatzhold2006a} contribute an energy that can arrest the collapse and stabilize the system as a finite sized droplet \cite{Saito2016a,Wachtler2016a,Bisset2016a}. Experiments have produced droplets by ramping a trapped condensate  into the dipole dominated regime leading to a single droplet or an array of droplets forming, depending on trap geometry \cite{Blakie2016a,Wachtler2016a,Bisset2016a}. Droplets with atom numbers in the range $10^3$--$10^4$ have been observed, with peak densities predicted to be an order of magnitude higher than the initial condensate density ($>\!10^{21}\mathrm{m}^{-3}$). The droplets are still well within the dilute weakly interacting regime, but three-body recombination becomes an important source of atom loss that limits droplet lifetime. Lifetimes of up to $\sim\!100\,$ms were measured for free-space droplets  \cite{Schmitt2016a}, with longer times observed for trapped droplets (e.g.~\cite{Kadau2016a}). The anisotropic DDI causes droplets to elongate along the direction that the dipoles are polarized into  highly anisotropic filaments.

 It is desirable to have a comprehensive understanding of the full excitation spectrum of the droplets. Indeed, in helium nanodroplets \cite{Dalfovo2001a}, which are dense self-bound superfluid droplets, the various types of bulk and surface excitations have been extensively studied for decades (e.g.~see \cite{Chin1995a,Stienkemeier2006a}).  Already some first steps have been made in dipolar droplets, with W{\"a}chtler \textit{et al.}~using a variational ansatz to characterize three shape oscillations \cite{Wachtler2016b}, with their prediction for the frequency of the axial mode comparing favorably to experiments with erbium \cite{Chomaz2016a}.  Here we present the results of the first calculations of the full excitation spectrum of a dipolar condensate in the self-binding regime by solving the Bogoliubov-de Gennes equations. We study the modes bound by the elongated droplet in free-space and the nature of instability as the number of atoms in the droplet decreases towards the critical number. Also, by including a trapping potential we quantify the evolution of the spectrum from a trap-bound condensate into a self-bound droplet.

\emph{Formalism}-- 
Several works \cite{Saito2016a,Ferrier-Barbut2016a,Wachtler2016a,Schmitt2016a,Wachtler2016b,Bisset2016a,Baillie2016b,Chomaz2016a,Boudjemaa2015a,Boudjemaa2016a,Boudjemaa2017a,Oldziejewski2016a,Macia2016a} have  established that the ground states and dynamics of a dipolar condensate in the droplet regime is well-described by a generalized nonlocal Gross-Pitaevskii equation (GPE). The time-independent version for the ground state wavefunction $\psi_0$ has the form $\mu\psi_0=\LGP\psi_0$, where $\mu$ is the chemical potential and
\begin{align}
    \LGP &\equiv  -\frac{\hbar^2\nabla^2}{2M}  +  \Phi(\bx) + \gammaQF |\psi_0|^3.\label{e:LGP}
\end{align} 
The effective potential $ \Phi(\bx)\!=\! \int d\bx' \,U(\bx\!-\!\bx')|\psi_0(\bx')|^2$ describes the two-body interactions where
\begin{align}
   U(\br) &= g_s\delta(\br) +\frac{3g_{dd}}{4\pi r^3} (1-3\cos^2 \theta).
\end{align}
Here $g_s=4\pi a_s \hbar^2/M$ is the $s$-wave coupling constant, $a_s$ is the $s$-wave scattering length, and $g_{dd}=4\pi \add \hbar^2/M$ is the DDI coupling constant, with $\add=M\mu_0\mu^2/12\pi\hbar^2$ the dipole length determined by the magnetic moment $\mu_m$ of the particles. The DDI term is for dipoles polarized along the $z$ axis, and $\theta$ is the angle between $\br$ and the $z$ axis. The leading-order LHY correction to the chemical potential is $\Delta \mu = \gammaQF n^{3/2}$, which is included in Eq.~\eqref{e:LGP} using the local density approximation $n \!\rightarrow \! |\psi_0(\bx)|^2$, with coefficient $\gammaQF\! =\! \frac{32}{3}g_s\sqrt{\frac{a_s^3}{\pi}}(1+\tfrac32 \edd^2)$ 
\cite{Lima2011a,Bisset2016a} where $\edd\equiv\add/a_s$ \footnote{The LHY term derived for the homogeneous system has a small imaginary part, which for our main case of $\edd = 1.64$ is $\gammaQF^\mathrm{homo} = \gammaQF(1 + 0.01i)$ \cite{Lima2011a}. This arises from unstable modes in the Bogoliubov treatment of a homogeneous condensate. Due to finite-size and LHY fluctuation effects, there are no unstable modes in the droplet and we neglect the imaginary part. Droplet lifetime will be limited by three-body loss as observed in experiments \cite{Chomaz2016a,Schmitt2016a}.}.

The collective excitations of this system are Bogoliubov quasiparticles, which can be obtained by linearizing the time-dependent GPE $i\hbar\dot{\psi}=\LGP\psi$ about the ground state as 
\begin{align}
\psi\!=\!e^{-i\mu t/\hbar}\left[\psi_0\!+\!\sum_\nu\left( \lambda_\nu u_{\nu}e^{-i\epsilon_\nu t/\hbar}-\lambda_\nu^*v_{\nu}^*e^{i\epsilon_\nu t/\hbar}\right)\right]\!\!,\! 
\end{align}
(e.g.~see \cite{Morgan1998a,Ronen2006a}),  where $\lambda_\nu$ is the perturbation amplitude.
The quasiparticle modes  ${u_\nu,v_\nu}$ and energies $\epsilon_\nu$ satisfy the Bogoliubov-de Gennes (BdG) equations
\begin{align}
    \!\!\!\begin{pmatrix}  \LGP - \mu + X & -X \\
      X &\! -(\LGP - \mu + X)\end{pmatrix}\!\begin{pmatrix} u_\nu \\ v_\nu\end{pmatrix}
     &= \epsilon_\nu \!\begin{pmatrix} u_\nu \\ v_\nu\end{pmatrix},
\end{align}
where $X$ is the exchange operator given by
\begin{align}
  Xf &\equiv\psi_0\!\int \!d\bx'U(\bx\!-\!\bx') f(\bx')\psi_0^*(\bx')
  \!+\! \tfrac32 \gammaQF |\psi_0|^3f.
\end{align}
We normalize the quasiparticles according to $\int d\mathbf{x}(|u_\nu|^2-|v_\nu|^2)=1$. Solving the GPE for $\psi_0$ and BdG equations for the excitations has to be done numerically. We utilize the cylindrically symmetry of the problem (e.g.~see \cite{Ronen2006a,Blakie2012a}) to solve independently for excitations in different $m$-subspaces, where $m$ is the $z$-projection of angular momentum and employ a cylindrical cutoff for the DDI \cite{Lu2010a}. 
\begin{figure}[htbp!] 
   \centering
      \includegraphics[width=3.0in]{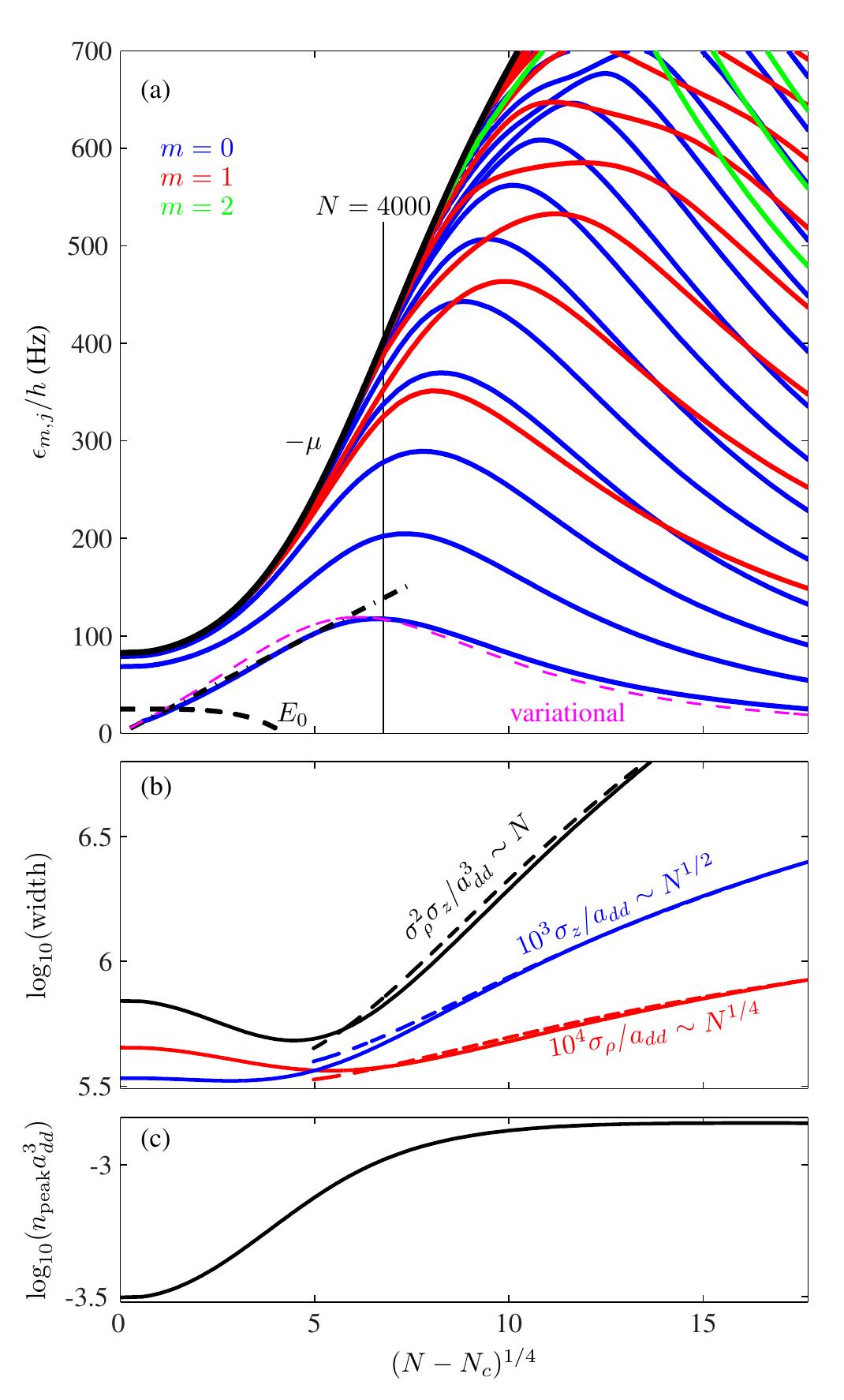}
   \caption{(a) Spectrum of a self-bound droplet of  $^{164}$Dy atoms with $a_s=80\,a_0$ in free-space as a function of $N$. The total droplet energy $E_0\!=\!\int\!d\mathbf{x}\psi_0^*(-\frac{\hbar^2\nabla^2}{2M}+\frac{1}{2}\Phi+\frac{2}{5}\gamma_{\mathrm{QF}}|\psi_0|^3)\psi_0$ (dashed black line) and  $-\mu$ (solid black line) are shown. A straight dash-dotted line fit shows that the lowest energy mode scales as $(N-N_c)^{1/4}$ as  $N$ approaches the critical number $N_c\approx1899$. Excitations are $m=0$ (blue), $m=1$ (red) and $m=2$ (green). Variational solution [with $N_c=2193$] for the lowest mode (magenta). (b) Widths $\sigma_\rho$, $\sigma_z$, and effective volume $\sigma_\rho^2\sigma_z$ (solid) as a function of $N$  and the corresponding large $N$ scaling (dashed). (c) Peak density as a function of $N$.   }
   \label{fig:FreeDropSpec}
\end{figure}

\emph{Self-bound droplets}--  
A universal phase diagram showing the conditions where self-bound solutions of the GPE exist was presented in Fig.~2 of Ref.~\cite{Baillie2016b}. This phase diagram only depends on the parameters $\edd$ and $N$, and shows that in the dipole dominated regime ($\edd>1$) there always exists a minimum critical number $N_c$ above which a stable droplet exists. As $\edd$ increases the critical number $N_c$ decreases.  
 Fig.~\ref{fig:FreeDropSpec}(a) shows the excitation spectrum as $N$ varies for  self-bound droplet of  $^{164}$Dy with $a_s=80\,a_0$ (i.e.~$\edd\approx1.63$). For this value of $\edd$ the critical atom number is $N_c\simeq1899$. 
 In solving for the excitations we find that they can be categorised into two types. (i) Those with $\epsilon_\nu<-\mu$  are bound by the droplet (noting that quasiparticle energies are relative to $\mu$, so these excitations have negative energy). (ii) Those with $\epsilon_\nu>-\mu$ are hence unbounded (part of the continuum) and are sensitive to the details of the finite numerical grid used in the calculations. We only show the bound excitations in Fig.~\ref{fig:FreeDropSpec}(a) and indicate $-\mu$ for reference. Our results show that the number of these bound excitations increases with $N$. As $N$ is reduced towards $N_c$ (where only a few of excitations remain) the lowest $m=0$ mode goes soft indicating the onset of a dynamical instability of the self-bound state. This mode softens as $\epsilon_{0}\sim(N-N_c)^{1/4}$ [see Fig.~\ref{fig:FreeDropSpec}(a)], similar to the behaviour predicted at the instability point of attractive condensates \cite{Ueda1998a} and droplets in binary condensates \cite{Petrov2015a}. 

We also observe that the energy of the lowest quasiparticle initially increases with $N$ until it reaches a maximum at  $N\approx4\times10^3$. This mode has a monopole (compressional) character for $N\lesssim4\times10^3$, and its softening indicates increasing system compressibility  as $N\to N_c$. For $N\gtrsim4\times10^3$ this mode exhibits a quadrupolar character, consistent with the system becoming incompressible (e.g. see \cite{Stringari1996a}).  For comparison we show the energy of this mode obtained by a variational Gaussian treatment \cite{Wachtler2016b,Yi2001a}, which we find to be a good description of the full numerical result. We note the variational theory predicts a 15\% higher value for $N_c$.
 
The droplet compressibility is also revealed directly from condensate properties. In Figs.~\ref{fig:FreeDropSpec}(b)-(c) we show the peak density $\npeak$, the widths \footnote{ $\sigma_\nu$ is the distance where the density falls to $\frac{1}{e}$ of $\npeak$ along the $\nu$-axis.} $\{\sigma_\rho,\sigma_z\}$, and the effective volume $\sigma_\rho^2\sigma_z$ of the condensate as $N$ varies. For $N\lesssim4\times10^3$ (compressible regime) the widths and effective volume decrease with increasing $N$ and the peak density increases. For  $N\gtrsim4\times10^3$ the system behaves like an incompressible liquid: the peak density $\npeak$ remains constant, and the widths scale so the volume changes linearly with $N$.

\begin{figure}[htbp] 
    \centering
      \includegraphics[trim=5 0 0 0,clip=true,width=3.5in]{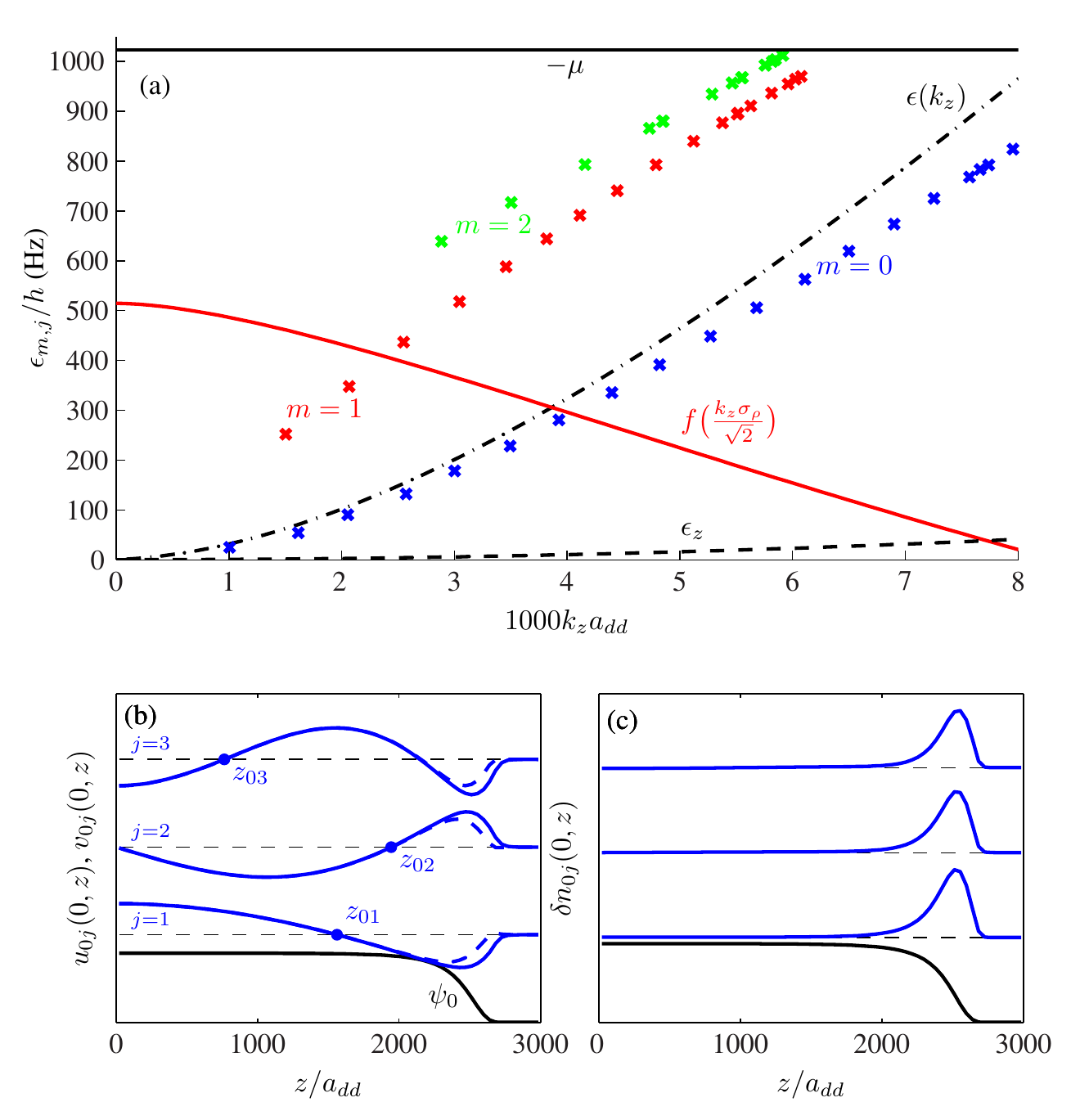}
      \caption{(a) Discrete dispersion relation for a self-bound droplet of $N=10^5$ $^{164}$Dy atoms with $a_s=80\,a_0$ in free-space. Also shown are $-\mu$ (solid black line), $\epsilon_z$ (dashed), $\epsilon(k_z)$ from Eq.~\eqref{e:eq1d} (dash-dotted) and $f$ with an arbitrary scale (red line). Excitations (crosses) are mapped using $k_z$ (see text). (b) Lowest three $m=0$ quasiparticles $u_{mj}$ (solid blue) $v_{mj}$ (dashed blue) modes, and (c) associated density fluctuations $\delta n_{mj}$ (see text). The quasiparticle results in (b) and (c) are vertically offset for clarity, and the condensate amplitude $\psi_0$ is shown for reference (black).}
   \label{fig:uvplot}
\end{figure}

In the incompressible region there are many bound modes that form a ladder of regularly spaced excitations [see Fig.~\ref{fig:FreeDropSpec}(a)]. The lowest energy mode, that we discussed above, is the first of the ladder of $m=0$ excitations.  At a higher energy a ladder of $m=1$ excitations begins, and so on (higher $m$-ladders) until the $-\mu$ threshold is crossed. These modes tend to be confined to the region of space occupied by the condensate. Noting that in the incompressible regime the condensate has the shape of a long filament [$\sigma_z\gg\sigma_\rho$, see Fig.~\ref{fig:FreeDropSpec}(b)], the ladder of modes corresponds to a sequence of harmonics along the $z$-extent of the condensate as shown for the lowest three modes in Fig.~\ref{fig:uvplot}(b). We see that the $u_\nu$ and $v_\nu$ quasiparticle amplitudes are essentially identical within the central  region of the condensate where the density is saturated. The density fluctuation associated with a quasiparticle is given by  $\delta n_\nu\sim (u_\nu-v_\nu)\psi_0$, and thus vanishes inside the condensate [see Fig.~\ref{fig:uvplot}(c)], consistent with the incompressible character of this regime (c.f.~\cite{Ronen2006a,Bisset2013a,Blakie2013a}). These results also show these excitations mainly perturbing the density in the surface region. The ``centrifugal potential'' for higher $m$ excitations shifts their ladders to higher energy. However as $N$ increases the filament width grows as $N^{1/4}$ [see Fig.~\ref{fig:FreeDropSpec}(b)] and higher $m$-ladders are increasingly bound within the droplet.

We can also quantify the character of the bound modes by assigning a wavevector to each quasiparticle to compute a discrete dispersion relation. 
We set $z_\nu$ as the first solution of $u_\nu(0,z)=0$ for $z>0$ and define the wavevector $k_z= \pi/2z_\nu$ ($k_z=\pi/z_\nu$) for even (odd) modes  [see Fig.~\ref{fig:uvplot}(b)]. 
The results of this analysis are shown in Fig.~\ref{fig:uvplot}(a), where the different ladders of $m$-excitations are clearly seen.   

The $m=0$ discrete dispersion relation is well described by the quasi-one-dimensional (quasi-1D) result found by assuming a Gaussian radial profile of the condensate and excitations with width $\sigma_\rho$  
\begin{align}
 \epsilon(k_z) &\!=\! 
 \sqrt{ \epsilon_z^2 + 2\epsilon_z \npeak \!\!\left[\frac{g_s}{2} \!-\!\frac{g_{dd}}{2}f\biggl(\!\frac{k_z\sigma_\rho}{\sqrt2}\!\biggr) \!+\! \frac35 \gammaQF \npeak^{1/2}  \right]},
 \label{e:eq1d}
\end{align}
where $\epsilon_z = \hbar^2k_z^2/2M$ and $f(q) = 1+3q^2e^{q^2}\Ei(-q^2)$ is the quasi-1D DDI \cite{Giovanazzi2004a}, with $\Ei$ being the exponential integral. The dispersion relation (\ref{e:eq1d}) only requires $\sigma_\rho$ and $\npeak$ from the GPE solution and has no other fitting parameters. There is no apparent linear (phonon) dispersion for the $m=0$ results in Fig.~\ref{fig:uvplot}(a). However, the kinetic energy of these modes is also negligible (the free-particle dispersion $\epsilon_z$ is shown for reference), and $ \epsilon(k_z)$ still accurately fits the $m=0$ modes if we neglect the $\epsilon_z^2$ term. Thus interaction effects (including the LHY term) dominate the energetics of these modes, with the curvature in the dispersion relation arising from the momentum dependence of the DDI, described by $f$.  The variation of $f$ with $k_z$ is set by the radial width $\sigma_\rho$, and we see that $f$ changes rapidly over the $k_z$ range spanned by the bound modes [see Fig.~\ref{fig:uvplot}(a)]. The dispersion relation \eqref{e:eq1d} also has some similarities to that of a dilute bilayer system of polar molecules with a three-body interaction \cite{Lu2015a}.

We have calculated spectra for free-space droplets over a wide parameter regime ($a_s/a_0\in\{70,80,90,100\}$ and $N\in \{N_c, \dots, 2\times 10^5\}$) and find the general spectrum behavior to be qualitatively similar to the results of this section.

\emph{Transition to self-bound droplets in a trap}-- 
Dipolar condensates are typically prepared by cooling the atoms through the condensation transition in an external trap with $\edd<1$. In this regime the role of quantum fluctuations is unimportant and the condensate profile is determined by a balance between the repulsive two-body interactions and the trapping potential (i.e.~exhibits a Thomas-Fermi density profile \cite{Eberlein2005a}). From this point a Feshbach resonance is used to reduce $a_s$ (i.e.~increase  $\edd$) to bring the system into the dipole dominated regime where droplets can form (e.g.~see \cite{Kadau2016a,Ferrier-Barbut2016a,Chomaz2016a,Schmitt2016a}). It is interesting to explore the nature of the excitation spectrum as the condensate undergoes the transition from being bound by the trapping potential to being self-bound as a droplet.
 Trap geometry can play a significant role in the stability properties and excitations of a dipolar condensate (e.g.~see \cite{Santos2003a,Ronen2006a,Ronen2007a,Blakie2012a,Bisset2013b}), however here we focus on the case of a spherically symmetric trap where the condensate smoothly crosses over into a droplet as $\edd$ increases. 
Our solutions for the condensate and excitations are found using the procedure outlined earlier with the potential $V_{\mathrm{trap}} =  \tfrac12  M \omega^2|\bx|^2$ 
added to $\LGP$.

\begin{figure}[htbp] 
    \centering
      \includegraphics[width=3.1in]{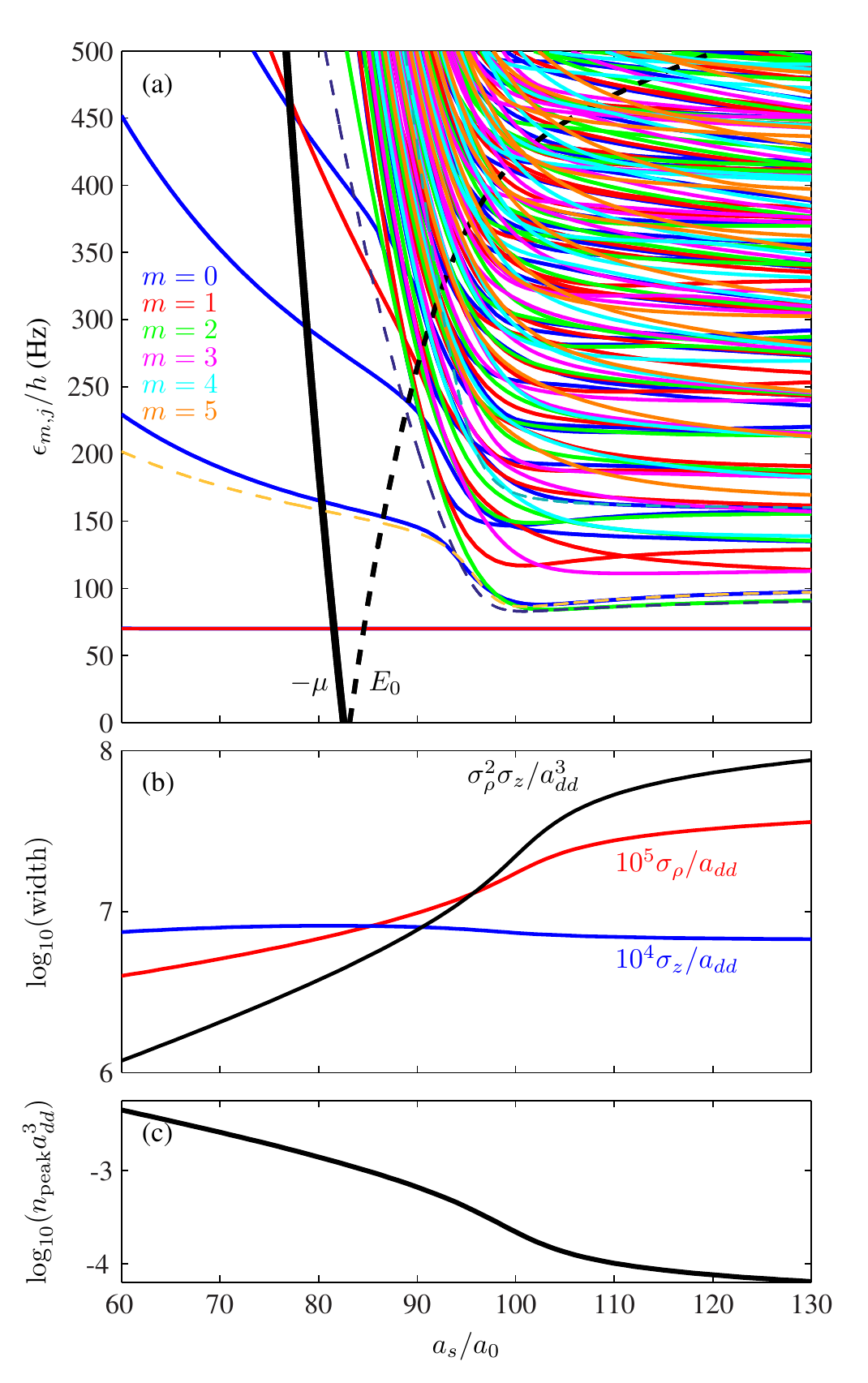}
      \caption{(a) Spectrum of $2\times10^4$ $^{164}$Dy atoms in a spherical trap with $\omega =2\pi\times70\:\mathrm{s}^{-1}$ showing modes up to $m=5$. The condensate energy (thick black dashed line) and $-\mu$ (heavy black line) are shown. The three oscillation modes predicted by variational theory (thin dashed lines). The Kohn modes at $\epsilon_\nu=\hbar\omega$ (note $m=1$ mode obscures the $m=0$ mode). (b) Widths $\sigma_\rho$, $\sigma_z$, and effective volume $\sigma_\rho^2\sigma_z$ (solid) as a function of $a_s$. (c) Peak density as a function of $a_s$. }
   \label{fig:spheretrap}
\end{figure}

Our results for a system of $2\times10^4$ $^{164}$Dy atoms are shown in Fig.~\ref{fig:spheretrap}.  For $a_s\gtrsim95a_0$ the condensate is in a low density trap bound state and has a dense excitation spectrum. As $a_s$ decreases below $95a_0$ the condensate energy rapidly decreases into negative values as the droplet self-binds. As this happens most of the quasiparticle energies start rapidly rising into a quasi-continuum of excitations that are bound by the trap but not within the droplet. A few $m=0$ modes and a single $m=1$ mode are seen to ``peel off'' from these rapidly rising modes and become bound excitations within the droplet, similar to those shown in Fig.~\ref{fig:FreeDropSpec} [the spectrum here at $a_s=80a_0$ is similar to that of the free-droplet in Fig.~\ref{fig:FreeDropSpec}(a) at $N=2\times10^4$].  We also indicate $-\mu$, but note it is only an approximate estimate of the energy scale for self-bound excitations in the trapped system. Since the quasiparticle energies are measured relative to $\mu$,  it is useful to consider $\epsilon_\nu+\mu$ (not shown), which shows that the rapidly rising states are instead approximately constant in energy relative to the confinement potential, while the bound state energies rapidly become negative as the droplet forms. 

We also give measures of the condensate size and peak density for reference in Figs.~\ref{fig:spheretrap}(b) and (c). This shows that as $a_s$ decreases over the range of values shown, the radial width of the condensate reduces by about an order of magnitude while the axial length slightly increases (it is being constrained by the trap), thus becoming an elongated droplet. Also the peak density increases by more than an order of magnitude.

Because of the harmonic trap the system has Kohn modes, corresponding to center-of-mass oscillations at the trap frequency \cite{KohnThrm}, which are degenerate for our case of a spherically symmetric trap. As the trap frequency is reduced to zero these Kohn modes vanish reflecting the translational symmetry of the free droplet.

The parameters of Fig.~\ref{fig:spheretrap} match those in Fig.~7 of Ref.~\cite{Wachtler2016b}, where the variational approximation of the shape oscillation modes was developed. Those three modes are reproduced in Fig.~\ref{fig:spheretrap}(a) for reference. We see that two of the variational modes are low lying excitations in the trap bound regime ($a_s\gtrsim95a_0$). These modes cross as the system transitions into the droplet, and only one (with monopole/quadrupole character discussed in the free droplet section) remains a low energy mode. The third shape mode lies in the dense spectrum of excitations on both sides of the crossover.

\emph{Conclusions and outlook}--
In this paper we have presented the first calculations for the full excitation spectrum of a self-bound dipolar droplet. We find that as $N$ increases the droplet crosses over from being compressible to  behaving like an incompressible liquid, revealed by the frequency and character of the lowest energy excitation. We also observe that low angular momentum quasiparticles are bound within the elongated droplet and propagate as axial phonon modes. While we present figures for $^{164}$Dy, all results are also valid for other atoms with simple scaling. The $s$-wave scattering lengths must be scaled by $\add$ and frequencies must be scaled by $M\add^2$. E.g., for $^{166}$Er  the vertical axis of Figs.~\ref{fig:FreeDropSpec}(a), \ref{fig:uvplot}(a) and \ref{fig:spheretrap}(a) must be multiplied by 3.9  and the horizontal axis of Fig.~\ref{fig:spheretrap} must be multiplied by 0.5. The case $a_s=80a_0$ considered in Figs.~\ref{fig:FreeDropSpec} and \ref{fig:uvplot} corresponds to $40a_0$ for $^{166}$Er.

In experiments the excitations have been excited by modifying the trap (or an external field) \cite{Chomaz2016a} (also see \cite{Bismut2010a}). Such an approach will not effectively couple to the many spatially varying modes we have discussed here, and new techniques will be needed  to  explored to measure these, 
such as  Bragg spectroscopy (e.g.~see \cite{Bismut2012a,Steinhauer2002a,Blakie2012a}) or extensions of this approach utilizing \emph{in situ} imaging \cite{Shammass2012a}.

\begin{acknowledgments}
DB and PBB acknowledge the contribution of NZ eScience Infrastructure (NeSI) high-performance computing facilities, and support from the Marsden Fund of the Royal Society of New Zealand.   
RMW acknowledges partial support from the National Science Foundation under Grant No.~PHY-1516421.  
\end{acknowledgments}

%

\end{document}